\documentclass[a4paper,11pt]{article}
\pdfoutput=1 

\usepackage{jcappub} 

\usepackage[T1]{fontenc} 

\title{\boldmath Directional Sensitivity of the NEWSdm Experiment to Cosmic Ray Boosted Dark Matter}

\author[a]{N.Y. Agafonova,}
\author[b,c]{A. Alexandrov,}
\author[d,1]{A.M. Anokhina\note{Corresponding author.},}
\author[b,c]{T. Asada,}
\author[a]{V.V Ashikhmin,}
\author[b,c]{V. Boccia,}
\author[b,e]{D. Centanni,}
\author[f]{M.M. Chernyavskii,}
\author[g]{N. D'Ambrosio,}
\author[b,c]{G. De Lellis,}
\author[b,c]{A. Di Crescenzo,}
\author[h]{Y.C. Dowdy,}
\author[i]{S. Dmitrievski,}
\author[a]{R.I. Enikeev,}
\author[j,k]{G. Galati,}
\author[d]{V.I. Galkin,}
\author[b]{A. Golovatiuk,}
\author[f]{S.A. Gorbunov,}
\author[i]{Y. Gornushkin,}
\author[l]{A.M. Guler,}
\author[d]{V.V. Gulyaeva,}
\author[b,c]{A. Iuliano,}
\author[d]{E.V. Khalikov,}
\author[m]{S.H. Kim,}
\author[f]{N.S. Konovalova,}
\author[n]{Y.O. Krasilnikova,}
\author[b,c]{A. Lauria,}
\author[m]{K.Y. Lee,}
\author[b,o]{V.P. Loschiavo,}
\author[d]{A.K. Managadze,}
\author[i]{A. Miloi,}
\author[b,c]{M.C. Montesi,}
\author[h,p]{T. Naka,}
\author[f]{N.M. Okateva,}
\author[m]{B.D Park,}
\author[d]{D.A. Podgrudkov,}
\author[f]{N.G. Polukhina,}
\author[d]{T.M. Roganova,}
\author[q]{G. Rosa,}
\author[d]{M.A. Samoilov,}
\author[o]{Z.T. Sadykov,}
\author[h]{K. Saeki,}
\author[p]{O. Sato,}
\author[a]{I.R. Shakiryanova,}
\author[f]{T.V. Shchedrina,}
\author[q]{T. Shiraishi,}
\author[m]{J.Y. Sohn,}
\author[i]{A. Sotnikov,}
\author[f]{N.I. Starkov,}
\author[f]{E.N. Starkova,}
\author[n]{D.M. Strekalina,}
\author[b]{V. Tioukov,}
\author[d]{E.D. Ursov,}
\author[n,o]{A. Ustyuzhanin,}
\author[i]{S. Vasina,}
\author[f]{R.A. Voronkov,}
\author[m]{C.S. Yoon}

\affiliation[a]{Institute for Nuclear Research of the Russian Academy of Sciences (INR RAS)\\117312, 60-letiya Oktyabrya Avenue 7a, Moscow, Russia}
\affiliation[b]{Sezione INFN di Napoli \\
Complesso universitario di Monte S. Angelo ed. 6 via Cintia, 80126, Napoli, Italy}
\affiliation[c]{Università di Napoli ``Federico II'' \\
CSI - Servizio PEC Complesso Universitario di Monte S. Angelo Via Cinthia - 80126 Napoli, Italy}
\affiliation[d]{Skobeltsyn Institute for Nuclear Physics Lomonosov Moscow State University (SINP MSU) \\
119991, Leninskie gory 1(2), Moscow, Russia}
\affiliation[e]{Università di Napoli Parthenope \\ Via Ammiraglio Ferdinando Acton, 38, 80133 Napoli NA, Italy}
\affiliation[f]{P.N. Lebedev Physical Institute (LPI RAS)\\
119991, Leninsky Anenue 53, Moscow, Russia}
\affiliation[g]{Laboratori Nazionali dell'INFN di Gran Sasso \\
Via G. Acitelli, 22 67100 Assergi L’Aquila, Italy}
\affiliation[h]{Toho University \\
2 Chome-2-1 Miyama, Funabashi, Chiba 274-8510, Japan}
\affiliation[i]{Joint Institute for Nuclear Research (JINR)\\ 141980, Joliot Curie Street 6, Dubna, Russia}
\affiliation[j]{Sezione INFN di Bari\\ Via E. Orabona 4, 70125, Bari, Italy}
\affiliation[k]{Università di Bari\\ Piazza Umberto I, 70121, Bari, Italy}
\affiliation[l]{Middle East Technical University (METU)\\ Üniversiteler Mahallesi, Dumlupınar Bulvarı No:106800 Çankaya Ankara, Turkey}
\affiliation[m]{Physics Education Department \& RINS, Gyeongsang National University\\ Jinju, Korea}
\affiliation[n]{National University of Science and Technology ``MISiS''\\
119049, Leninsky Avenue 4, Moscow, Russia}
\affiliation[o]{National Research University Higher School of Economics\\109028, Pokrovsky Bulvar 11, Moscow, Russia}
\affiliation[p]{Nagoya University\\ Furo-cho, Chikusa-ku, Nagoya, 464-8601, Japan}
\affiliation[q]{Sezione INFN di Roma\\ Piazzale Aldo Moro, 2 - 00185 Rome RM, Italy}
\affiliation[r]{Kanagawa University\\ 3-27-1 Rokkakubashi, Kanagawa-ku, Yokohama-shi, Kanagawa, 221-8686, Japan}

\emailAdd{anokhannamsu@gmail.com}

\abstract{We present a study of a directional search for Dark Matter boosted forward when scattered by cosmic-ray nuclei, using a module of the NEWSdm experiment. The boosted Dark Matter flux at the edge of the Earth's atmosphere is expected to be pointing to the Galactic Center, with a flux 15 to 20 times larger than in the transverse direction.

The module of the NEWSdm experiment consists of a 10 kg stack of Nano Imaging Trackers, i.e.~newly developed nuclear emulsions with AgBr crystal sizes down to a few tens of nanometers. The module is installed on an equatorial telescope. 
The relatively long recoil tracks induced by boosted Dark Matter, combined with the nanometric granularity of the emulsion, result in an extremely low background.
This makes an installation at the INFN Gran Sasso laboratory, both on the surface and underground, viable. A comparison between the two locations is made. 
The angular distribution of nuclear recoils induced by boosted Dark Matter in the emulsion films at the surface laboratory is expected to show an excess with a factor of 3.5 in the direction of the Galactic Center. This excess allows for a Dark Matter search with directional sensitivity.
The surface laboratory configuration prevents the deterioration of the signal in the rock overburden and it emerges as the most powerful approach for a directional observation of boosted Dark Matter with high sensitivity. 
We show that, with this approach, a 10 kg module of the NEWSdm experiment exposed for one year at the Gran Sasso surface laboratory can probe Dark Matter masses between 1 keV/c$^2$ and 1~GeV/c$^2$ and cross-section values down to $10^{-30}$~cm$^2$ with a directional sensitive search. }

\begin{document}
\maketitle
\flushbottom

\section{Introduction}
\label{sec:intro}
The NEWSdm (Nuclear Emulsion for WIMP Search with a directional measurement) detector was proposed~\cite{NEWS2016b} for a directional observation of target nuclei recoils induced by the scattering of Weekly Interacting Massive Particles (WIMP), as possible constituents of Dark Matter (DM). In NEWSdm, recoil tracks are visualized by a novel type of nuclear emulsion films, the NIT (Nano Imaging Trackers)~\cite{NIT-UNIT} where AgBr crystals have diameters down to 20~nm. 
The detector consists of a stack of emulsion films placed on an equatorial telescope to compensate for the Earth's rotation and to be sensitive to the apparent incoming direction of DM particles, through the detection of the induced nuclear recoil tracks.

The non-relativistic nature of dark matter induces low-energy nuclear recoils, with typical energy below 100 keV. In general, their detection requires extra pure detectors located underground in order to suppress the cosmic-ray induced background. This is also true for a directional-sensitive search with an emulsion-based detector where the typical track length is well below one micron. 

Sub-GeV dark matter (DM) candidates are of increasing interest, since long-favoured candidates at higher masses have not been detected so far. For a low-mass dark matter candidate, current model-independent constraints are very weak. The expected existence of a subdominant, but highly energetic, DM component generated through the collisions with cosmic rays paves the way for new searches~\cite{Pospelov2019}. This DM component is the so-called Cosmic-Ray-accelerated Dark Matter or CRDM.
Several studies of CRDM have been carried out as reported in~\cite{Cappiello,ChenXia,PROSPECT,KeikoNEWSdm}. Some estimates from the point of view of CRDM directional detection were made in~\cite{KeikoNEWSdm}. 

In this work, we present the sensitivity to CRDM of a module for the NEWSdm experiment consisting of a 10 kg NIT detector installed on an equatorial telescope. The aim is to observe nuclear 
(mostly Hydrogen, Carbon, Nitrogen and Oxygen) recoil tracks induced by the collision of a light DM candidate with the emulsion, having directional sensitivity. 
 
There are several ways in which high-energy cosmic rays can accelerate the original DM particles in our Galaxy. Here, we consider the elastic scattering of sub-GeV mass DM particles off the galactic cosmic-ray nuclei, with dark matter interaction cross-sections $\sigma_{\chi p}$ below $ 10^{-28}$~cm$^2$. Fluxes and energy distributions of CR nuclei have been obtained using the GALPROP v.57 software package~\cite{Galprop57}. All the nuclei from H to Fe have been considered, thus covering more than 95\% of the total flux of CR. Masses of DM particles are assumed to range from 1~keV/c$^2$ to 1~GeV/c$^2$. We account for the attenuation of the DM flux in the atmosphere, down to the level of the Gran Sasso surface laboratory, and in the rock above the detector in the case of underground installation. 

We consider the INFN Gran Sasso laboratory as the detector location and we compare the performance of the detector in two cases: installation at the ground level or in the underground laboratory. The exposure on the surface is possible thanks to the relatively long track lengths which correspond to a smaller background contamination than the one expected in the search for non-relativistic dark matter particles. The recent measurement of sub-MeV energy neutrons on the surface laboratory with a NIT detector has demonstrated this concept~\cite{Shiraishi-n}. 
This work shows that it is possible to carry out a highly sensitive directional search of CRDM in the direction pointing to the Galactic Center, where the highest concentration of DM particles and CR is expected.

This paper is organized as follows. Section~\ref{sec:two} describes the modeling scheme of DM spectra at the top of the Earth's atmosphere, at the LNGS surface laboratory level, and at the underground laboratory level, which requires a propagation through 1400~m of rock, equivalent to 3600~m of water. We have accounted for the attenuation of the DM flux in the atmosphere and in the rock. In section~\ref{sec:three}, we analyze the features of the induced nuclear recoil tracks. 
 
In Section~\ref{sec:nuclear_angle}, we introduce a technique for directional observations and report the angular distributions of the recoil tracks. The capability of distinguishing a signal excess in the direction of the Galactic Center is also considered in section~\ref{sec:nuclear_angle}. 
 The sensitivity of a 10 kg module of the NEWSdm experiment is reported in Section~\ref{sec:directional}, where we also compare the signal from the Galactic Center (GC) to the one from the perpendicular direction (GLat) within the plane of the Galaxy.

\section{CR-boosted Dark Matter particles production and spectra}
\label{sec:two}

The model of high-energy dark matter particles assumes the energy transfer via elastic collisions with cosmic-ray nuclei. To calculate the spectra of accelerated DM particles near the Earth we have used the GALPROP v.57 software package, providing the spectra of various cosmic-ray nuclei at different points of the Galaxy $d\Phi_i(\boldmath{r}, z)/dT_i$, where $\Phi_i$ is the flux of the $i$-th CR nucleus (number of particles per unit area, time and solid angle), $T_i$ --- its kinetic energy, $x,y,z$ --- the Cartesian coordinates of the Galactic system, $kpc$, $r=\sqrt{x^2+y^2}$.

We considered the contribution from 9 CR nuclei ($^1$H, $^4$He, $^{12}$C, $^{16}$O, $^{20}$Ne, $^{24}$Mg, $^{28}$Si, $^{32}$S and $^{56}$Fe), providing 95\% of the total flux.
 
\begin{figure}[t]
 \centering
 \includegraphics[width=0.9\textwidth]{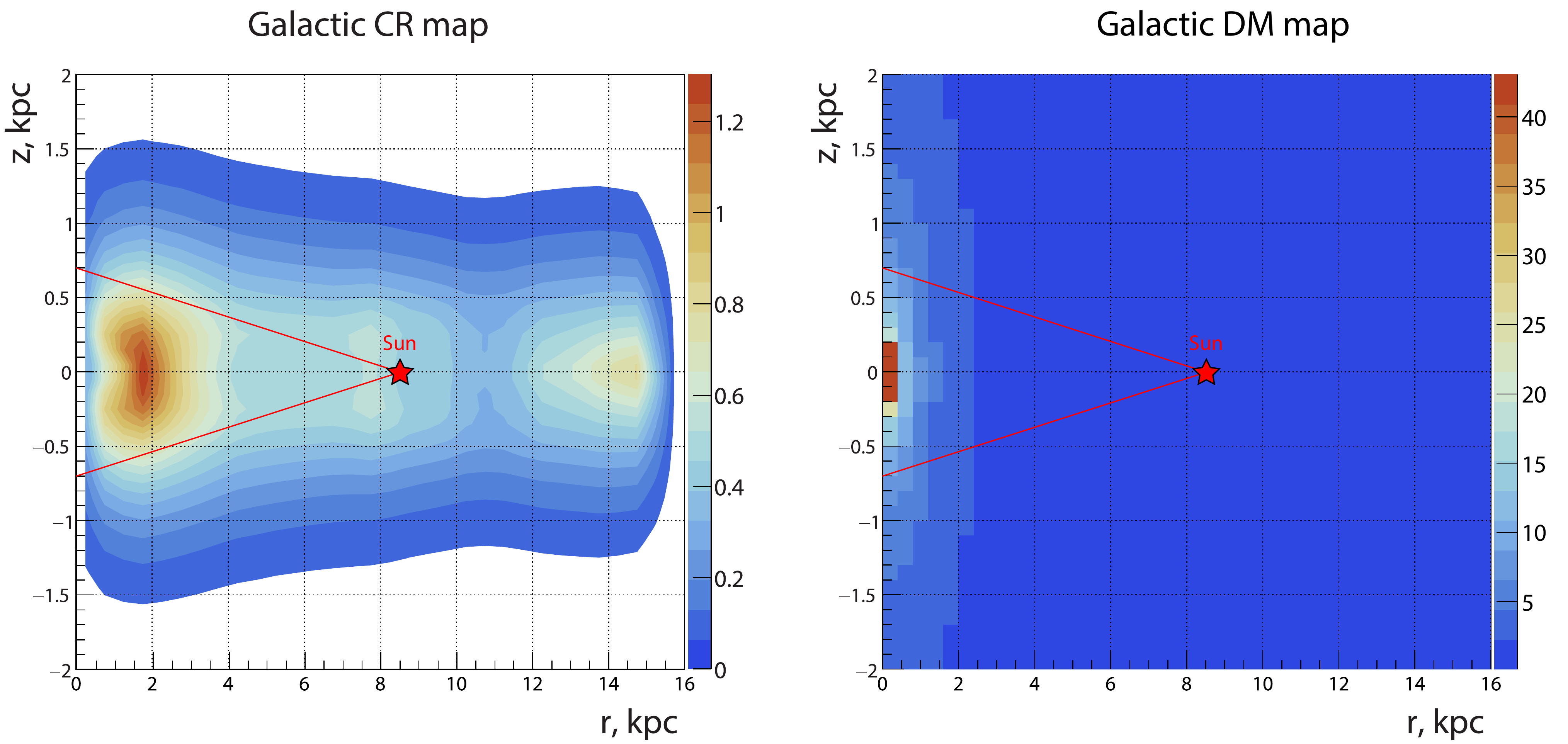} 
 \caption{Galactic Cosmic-Ray (by GALPROP v.57) and Dark-Matter (NFW) distributions in arbitrary units. The region within $\pm5$ degrees from the direction of the Galactic Center is also shown. $r$ denotes the distance from the Galactic Center in the galactic plane, $z$ corresponds to the direction perpendicular to the Galactic plane. 
}
 \label{pic2}
\end{figure}

Figure~\ref{pic2} depicts the maps of the cosmic-ray fluxes according to GALPROP v.57 and the dark matter distribution according to the Navarro-Frenk-White model (NFW)~\cite{NFW} $\rho_{\chi}^{NFS}(r_c)=\rho_s/[(r_c/r_s)(1+r_c/r_s)^2]$, where $r_s = 20$~kpc, $\rho_s=0.35$~GeV~cm$^{-3}$, $r_c = \sqrt{x^2+y^2+z^2}$.

\begin{figure}[t]
\centering
\includegraphics[width=0.7\textwidth]{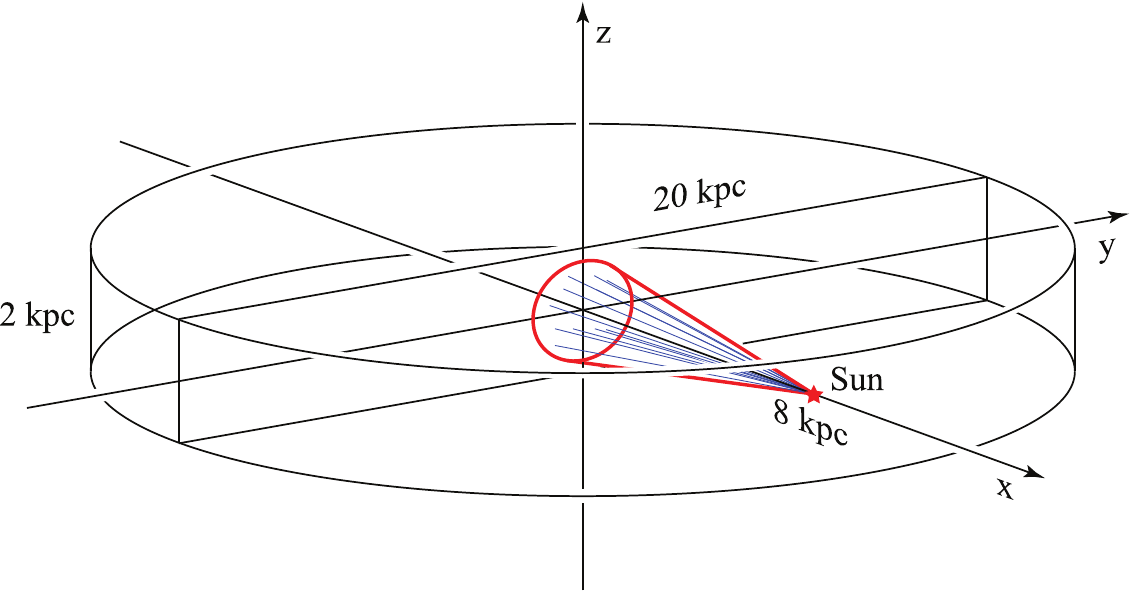} 
\caption{The averaging of boosted DM particle flux.} 
\label{pic3}
\end{figure}

The CR flux was averaged over a cone with an opening angle of 10 degrees aimed at the GC and with the vertex on the Sun (Earth). Lines of sight (l.o.s.) were originated by points uniformly distributed over the base of the cone, around the GC, and ended in the Solar system 8.5 kpcs away from the GC. 100 l.o.s. were considered as shown in figure~\ref{pic3}.

To calculate the fluxes of accelerated DM particles of different masses $m_{\chi}$ ranging from 1 keV/c$^2$ to 1 GeV/c$^2$, their single elastic collisions with CR nuclei inside a $\pm 5$ degree cone from the Sun to the GC have been considered (see figure~\ref{pic3}). 

The differential flux of CRDM is given by:
\begin{equation}
\frac{d\Phi_{\chi}}{dT_{\chi}}=\displaystyle\sum_{i} \int\limits_{l.o.s} dl \frac{\rho_{\chi}(\boldmath{r})}{m_\chi} \int\limits_{T_i^{min}} dT_i \frac{d\sigma_{\chi i}}{dT_{\chi}} \frac{d\Phi_i(\boldmath{r}, z)}{dT_i},
\label{eq1}
\end{equation}
where the sum is over all the considered CR nuclei.

The DM particle scattering cross-section off of an $i$-th CR nucleus (as used in~\cite{Pospelov2019,ChenXia,Cappiello}) is given by 
\begin{equation}
 \frac{d\sigma_{\chi i}}{dT_{\chi}}= \frac{F^2_i(q^2)A^2_i \mu^2_{\chi i}}{T^{max}_{\chi}(T_i) \mu^2_{\chi p}} \; \sigma_{\chi p}, 
 \label{eq2}
\end{equation}
where $F_i(q^2)$ is the form factor of the $i$-th CR nucleus, $A_i$ is the atomic mass number of the $i$-th nucleus, $\sigma_{\chi p}$ is the given DM particle-proton cross-section, $\mu_{\chi i}$ and $\mu_{\chi p}$ are the reduced masses for a DM particle interaction with the $i$-th CR nucleus and the proton, respectively. Just like in ~\cite{Pospelov2019,ChenXia} for H and He nuclei we used dipole form factor, for heavier nuclei we adopt the widely accepted Helm form factor parametrization~\cite{HelmFF}.

In the present work, only single tracks of recoil nuclei from elastic interaction with the emulsion are analyzed. The authors plan to consider the contribution of the quasi-elastic and deep inelastic interactions for the high-energy part of the CRDM spectrum in a future work.

\begin{figure}[ht]
\centering
\includegraphics[width=0.7\textwidth]{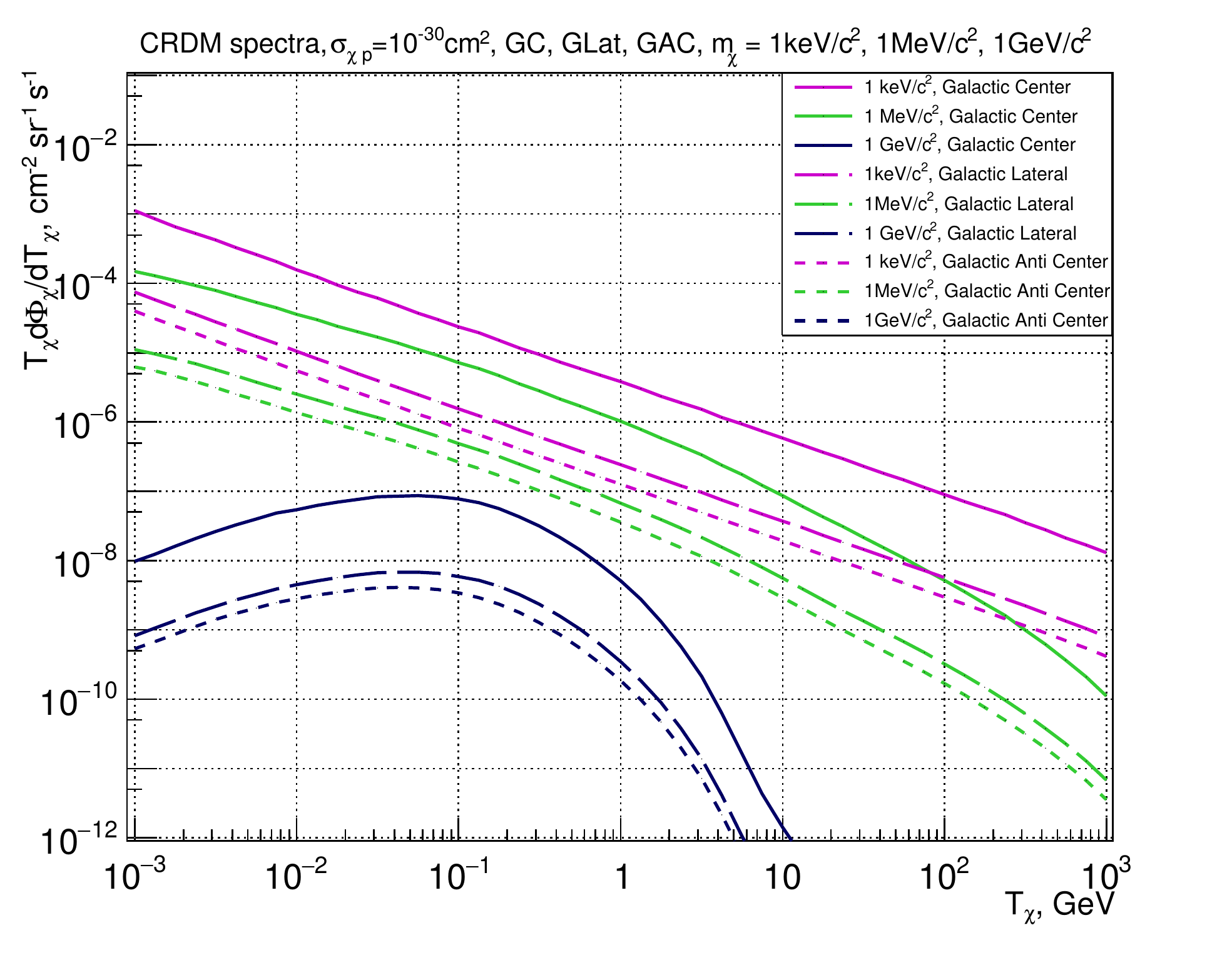}
\caption{CRDM spectra for $\sigma_{\chi p}=10^ {-30} $~cm$^2$ from Center, Anti Center of the Galaxy and from the Lateral direction.
Three DM masses (1 keV/c$^2$, 1 MeV/c$^2$, 1 GeV/c$^2$) are considered here as an example.}
\label{pic4}
\end{figure}
By integrating over the l.o.s. $l$ and the transferred energy, accounting for the contribution of the CR nuclei (see expression~\ref{eq1}), and averaging the signal over a cone with a $10$-degree aperture, we have derived the spectra of accelerated DM particles of different masses near the Earth, at the edge of the upper atmosphere, assuming the cross-section values of DM particle-proton interaction $\sigma_{\chi p} = 10^{-30}$ (see figure~\ref{pic4}). 

The CRDM spectra in figure~\ref{pic4} have a universal form and for $\sigma_{\chi p} = 10^{-29}$ and $10^{-28}$~cm$^2$ these spectra must be normalized using the coefficients 10 and 100 respectively.

Then, for each given cross-section value $\sigma_{\chi p}$ and for each DM mass, a sample of $10^{5}$ DM particles was generated according to the spectra reported in figure~\ref{pic4}. For comparison, figure~\ref{pic4} shows the spectra of CRDM in the direction of the Center, Anti Center of the Galaxy and from the perpendicular (Lateral) direction. These spectra were used to analyze the possibility of a directional signal search based on a comparison of the number of recoil nuclei tracks in the emulsion detector coming from the Center of the Galaxy and from other directions.

As a next step, the simulation of the propagation for each particle was carried out through the atmosphere (simplified as containing only nitrogen) to the altitude level of the Assergi village, where the surface INFN Gran Sasso Laboratories (LNGS) are located, and then through a layer of rock to the level of the Underground Facilities, in accordance with the cross-section $\sigma_{\chi p}$ (see figure~\ref{pic4}). The rock matter composition was simplified, so it only consisted of atoms with an average atomic mass $A=24$ a.m.u. and a density 
$\rho=2.7$~g/cm$^3$.

The kinematics of DM particle interactions with the target nucleus in the atmosphere, rock, and emulsion films in the NEWSdm experiment is given by the expressions~\ref{eq3},~\ref{eq4}, which describe the energy and momentum conservation laws for elastic interaction of relativistic CRDM. 
\begin{equation}
E_B=m_B[(E_A+m_B)^2+p_A^2\cos^2\theta_B]/[(E_A+m_B)^2-p_A^2\cos^2\theta_B],
\label{eq3}
\end{equation}
\begin{equation}
p_B=2m_Bp_A(E_A+m_B)\cos\theta_B/[(E_A+m_B)^2-p_A^2\cos^2\theta_B].
\label{eq4}
\end{equation}

 Particle $A$ denotes an accelerated DM particle, particle $B$ is the resting target nucleus before the interaction, $E_A$ and $p_A$ --- the energy and momentum of the DM particle before the collision, $E_B$ and $p_B$ --- nuclear recoil energy and momentum after the collision, $\theta_B$ --- the recoil angle.

The scattering model also includes the form factor of the nucleus, off which a DM particle scatters; for this study, we used dipole form factor~\cite{Pospelov2019, ChenXia} for hydrogen and Helm form factor~\cite{HelmFF, ChenXia} for heavier nuclei.

\begin{figure}[h]
 \centering
 \includegraphics[width=1.0\textwidth]{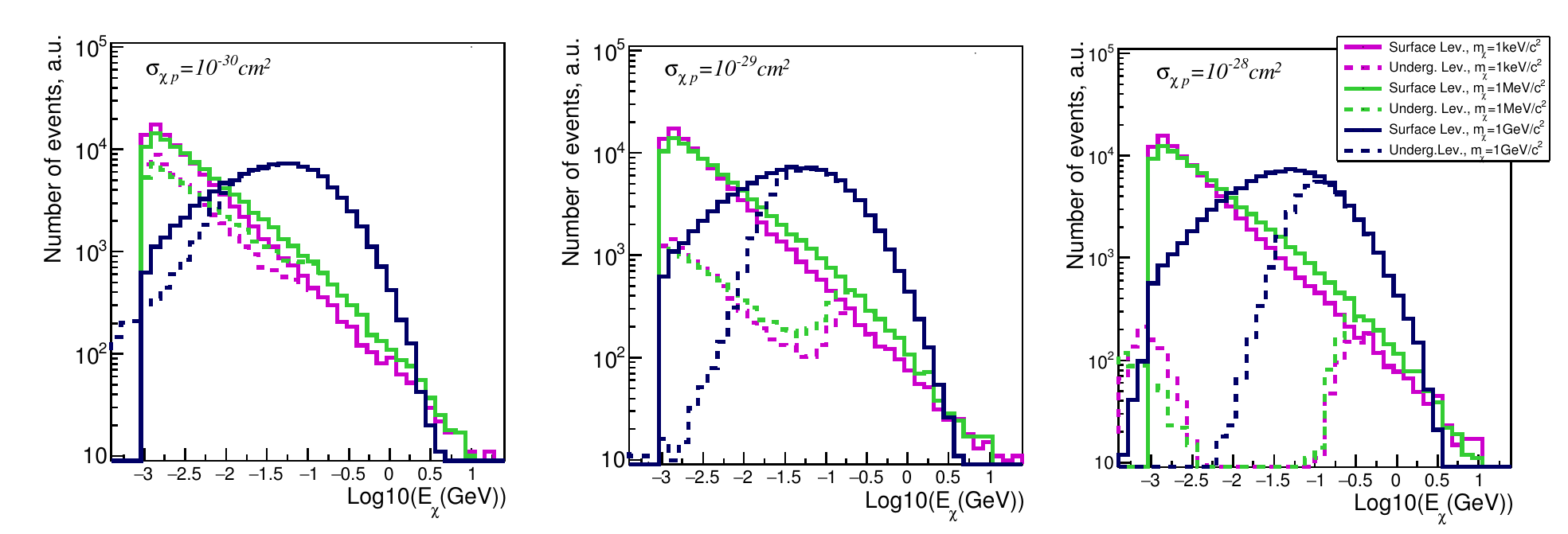} 
 \caption{Attenuation effect in the ground for $10^5$ DM events at the edge of the atmosphere for $\sigma_{\chi p} = 10^{-30}$, $\sigma_{\chi p} = 10^{-29}$, $\sigma_{\chi p} = 10^{-28}$, $m_{\chi}$ = 1 keV/c$^2$, 1 MeV/c$^2$ and 1 GeV/c$^2$. Solid lines denote the surface LNGS level, while dashed lines --- the underground LNGS level.}
 \label{fig-atten-G-2}
\end{figure}

For the detector material, i.e. nuclear emulsion, we have considered the scattering of DM particles on hydrogen and carbon nuclei (as a CNO group representative). Hydrogen, CNO group, Ag and Br nuclei are the main components of the emulsion. Heavy Ag and Br nuclei do not contribute significantly to the production of tracks with detectable length. The interaction of DM particles with H nuclei was considered also with regard to the spin-dependent part~\cite{Spin},~\cite{Spin2}, while DM scattering on carbon nuclei does not depend on spin since $J_{C}=0$.

Figure~\ref{fig-atten-G-2} shows the attenuation of the DM spectra after 1400 m of rock. The propagation through the atmosphere does not produce any sizeable effect on the DM spectra. All values of DM particle masses and three values of cross-sections ($\sigma_{\chi p}:$ $10^{-30}$, $10^{-29}$ and $10^{-28}$~cm$^2$) have been considered. As expected, significant differences in spectra at the surface and at the level of the Underground Lab are observed for the larger cross-section value. It is worth noting that the attenuation of the spectra of DM particles with different masses proceeds in different ways, therefore, the calibration of the experimental data obtained in underground observations faces a severe obstacle because the DM particle mass is not known.

The effect of attenuation has been modelled in several studies, ranging from a simple analytic approximation in~\cite{Pospelov2019} to Monte Carlo simulations in, e.g.,~\cite{ChenXia}, and has been shown in~\cite{atten-Xenon} to produce diurnal modulation of the signal.
In ~\cite{atten-Xenon} It was shown that the diurnal variation of the signal intensity in a liquid xenon detector located at the depth of 2~km underground can be useful for suppressing the background and increasing the sensitivity to sub-GeV accelerated dark matter.

The search for a signal above the neutrino background in the form of proton nuclear recoils was carried out by analyzing the Super-Kamiokande data~\cite{atten-S-K} with the detector located 1 km below ground level. Upper limits were set on the $\sigma_{\chi p}$ cross section from $10^{-33}$~cm$^{2}$ to $10^{-27}$~cm$^{2}$ for the DM mass range $m_{\chi}$ from 1 to 300 MeV/c$^{2}$. However, as noted in~\cite{atten-S-K}, the Cherenkov threshold for Super-Kamiokande is 0.5~GeV, so the energy range of nuclear recoils from tens of keV to hundreds of MeV has not been studied. It should also be noted that the scattering of DM particles on the nuclei of the Earth's soil leads to the fact that the nuclear recoils in the detector material deep underground (1 km for Super-Kamiokande) do not maintain the same direction of the DM particle source, i.e., the Galactic Center. 

Searches for Cosmic-Ray Boosted Sub-GeV Dark Matter were also carried out by the PandaX-II Experiment~\cite{atten-PANDA}.
The absence of candidates in the search for diurnal signal modulation led to the exclusion of a region of the parameter space corresponding to cross-section values from $10^{-31} $~cm$^{2}$ to $10^{-28} $~cm$^{2}$ and DM masses in the 0.1--300~MeV/c$^2$ range. However, the PandaX-4T detector is located at a depth of 6720~m.w.e., with a large expected attenuation and distortion of the energy spectra in the rock.

All of the circumstances mentioned above support a search using a detector located on the surface.

\section{Recoil tracks at the Surface and Underground LNGS laboratories}
\label{sec:three}

This section reports on the energy and angular distributions of H and C nuclear recoils at the surface LNGS Lab level and at the underground level at the point of the DM interaction.

Angular distribution of nuclear recoils can be described by a $\theta$ angle, which denotes the direction of the recoiling nucleus with respect to the direction of the Galactic Center. Angular distributions of nuclear recoils are relevant to estimate the sensitivity of a directional observation.

\begin{figure}[h]
\centering
\includegraphics{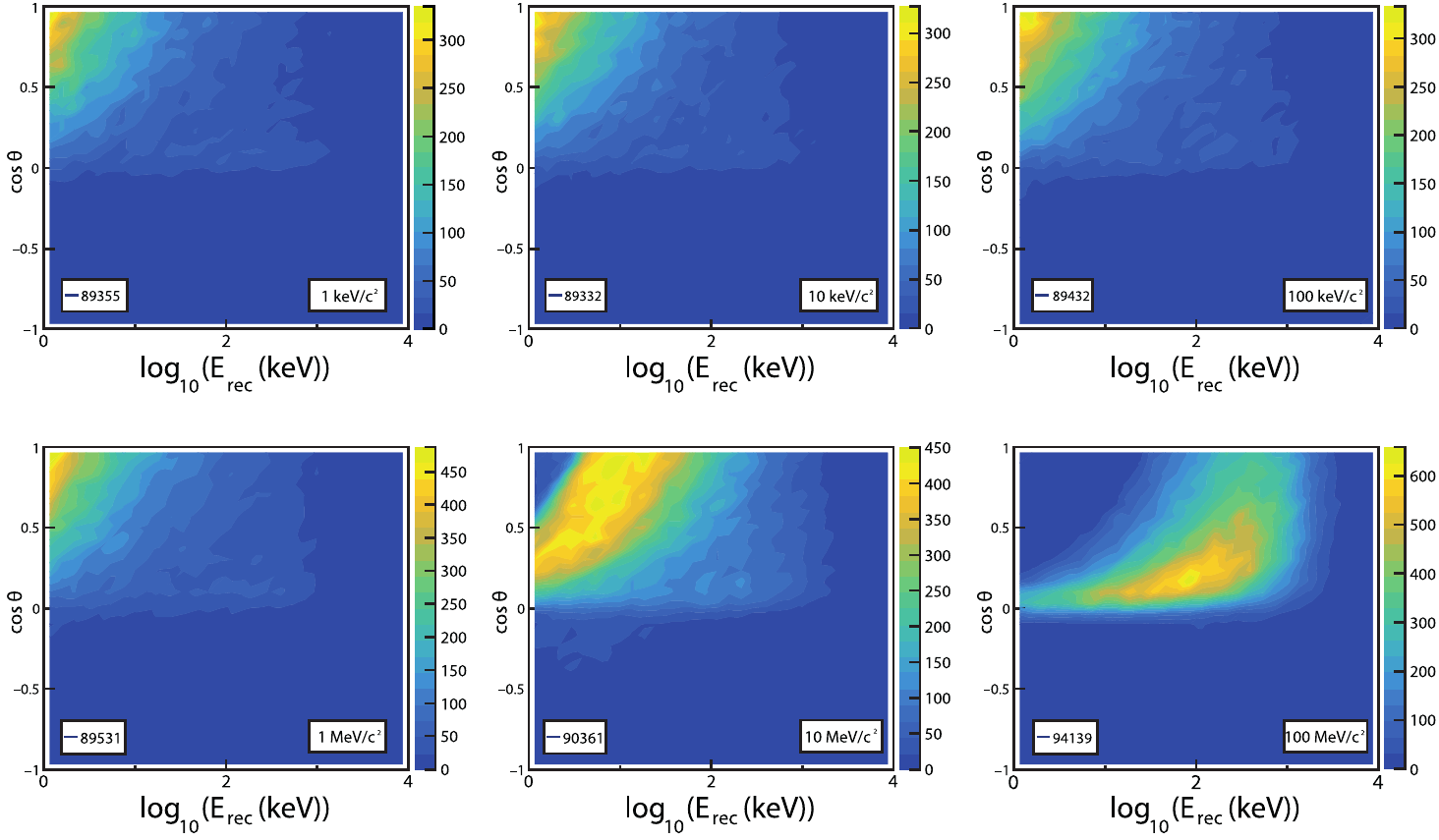}
\caption{Scatter plot of energy versus $\cos\theta$ for C recoils at the surface level for $\sigma_{\chi p}=10^{-28}$ cm$^2$.
The numbers in the figures denote the amounts of DM particles crossing the observational surface level for the first time, provided that the number of DM particles starting from the top of the atmosphere is set to $10^5$.}
\label{pic9}
\end{figure}

When calculating the sensitivity of the detector, it is important to determine the number of events at every step of the simulation. For our calculation, in every figure we provide the number of interactions for $10^{5}$ primary particles at the edge of the atmosphere. Two types of nuclei are considered: hydrogen and carbon. Figure~\ref{pic9} shows the examples of two-dimensional carbon nuclear recoil distributions at the surface level for $\sigma_{\chi p}=10^{-28} $~cm$^2$. It is important to note the wide angular distribution of C nuclear recoils and the ``energy-$\cos \theta$'' correlation at the surface level: the higher the carbon nuclear recoil energy, the better they retain the directionality to the source of primary DM particles.

\begin{figure}[h]
\centering
\includegraphics[width=0.98\textwidth]{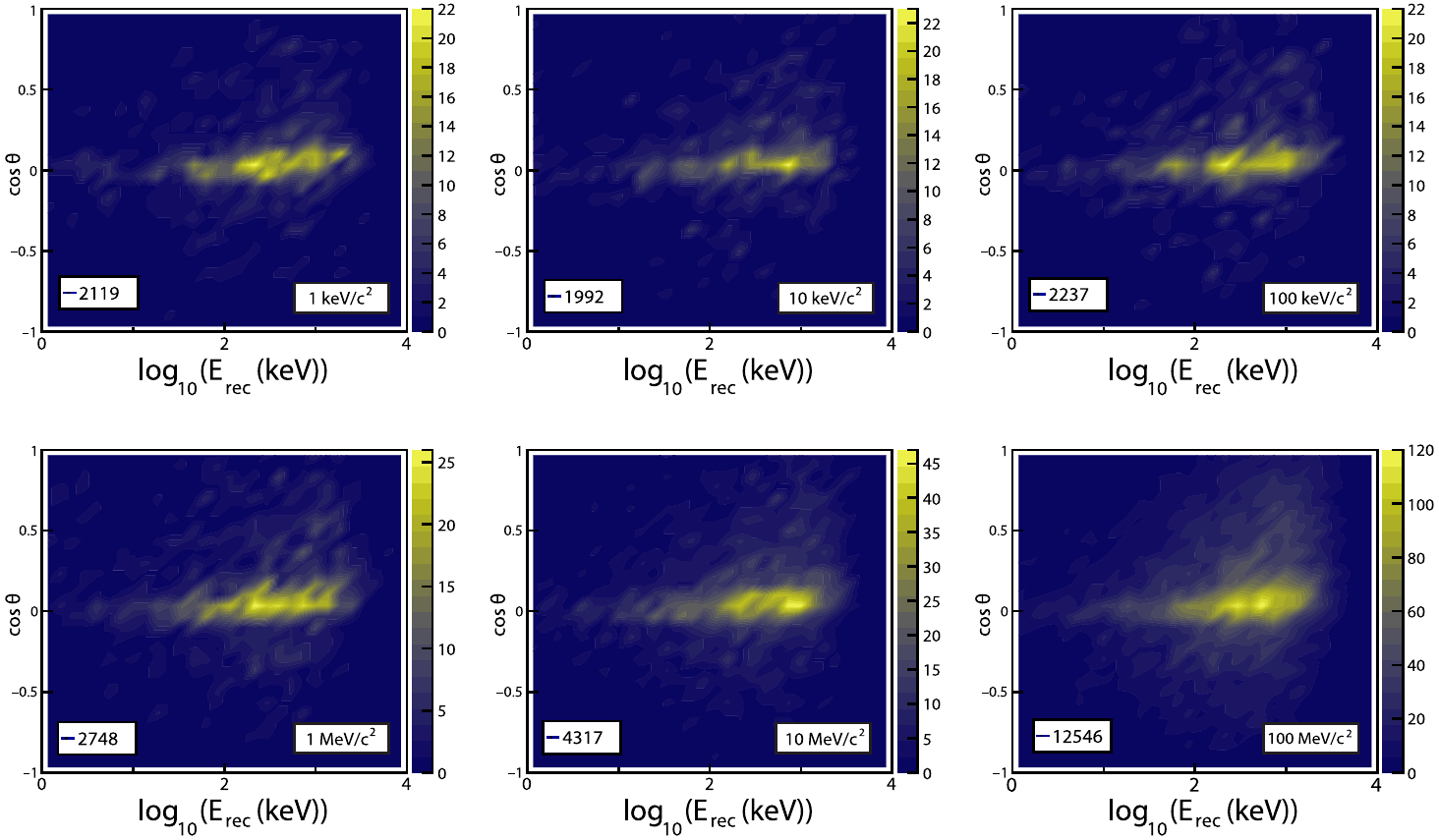}
\caption{Scatter plot of energy versus $\cos\theta$ for C recoils at the underground level for $\sigma_{\chi p}=10^{-28} $~cm$^2$.
The numbers in the figures denote the amounts of DM particles crossing the observational underground level for the first time, provided that the number of DM particles starting from the top of the atmosphere is set to $10^5$.}
\label{pic9u}
\end{figure}

The flux of DM particles at the underground laboratory is attenuated and their energy and angular spectra are distorted, thus affecting the directional observation, see figure~\ref{pic9u}.

In order to estimate the sensitivity of the emulsion detector to DM particles, a track length threshold for the event detection needs to be introduced. A track length threshold of 70~nm can be derived from the study of both the track formation processes from elementary emulsion structures (the so-called microclusters or grains) and the technical capabilities of image analysis procedures with track direction sensitivity~\cite{Andrey_plasmon1,Andrey_plasmon2}. 

A recent paper~\cite{Shiraishi-n} on the sub-MeV neutron measurement at the Gran Sasso surface laboratory with NIT emulsion films has shown that a background-free measurement can be carried out for track lengths down to the scale of 1 micron. Laboratory measurements have shown that to further reduce this threshold while keeping the background negligible, a dedicated emulsion production in a radon-free environment is needed. This is now being pursued at LNGS. Therefore, we assume that a background-free search can be carried out down to the 70 nm threshold level in this paper.

Modeling of the nuclear recoil tracks in the detector material has been done using the GEANT4~\cite{Geant1,Geant2,Geant3} toolkit with StandardNR --- Nuclear Recoil Physics List. Model of mass-fraction composition of nuclear emulsion is taken from~\cite{NIT-UNIT} (Ag --- 44.5, Br --- 31.7, I --- 2.0, C --- 10.1, H --- 1.6, O --- 7.4, N --- 2.7 \%), with density $\rho$ = 3.44~g/cm$^3$. Since the number of carbon nuclei in the emulsion is approximately the same as oxygen and nitrogen nuclei altogether, the signal from the entire CNO group is obtained as twice as much of the one from the carbon nuclei. Track lengths are defined as the length of the straight line best fitting the track points.

\begin{figure}[h]
\centering
\includegraphics[width=0.9\textwidth]{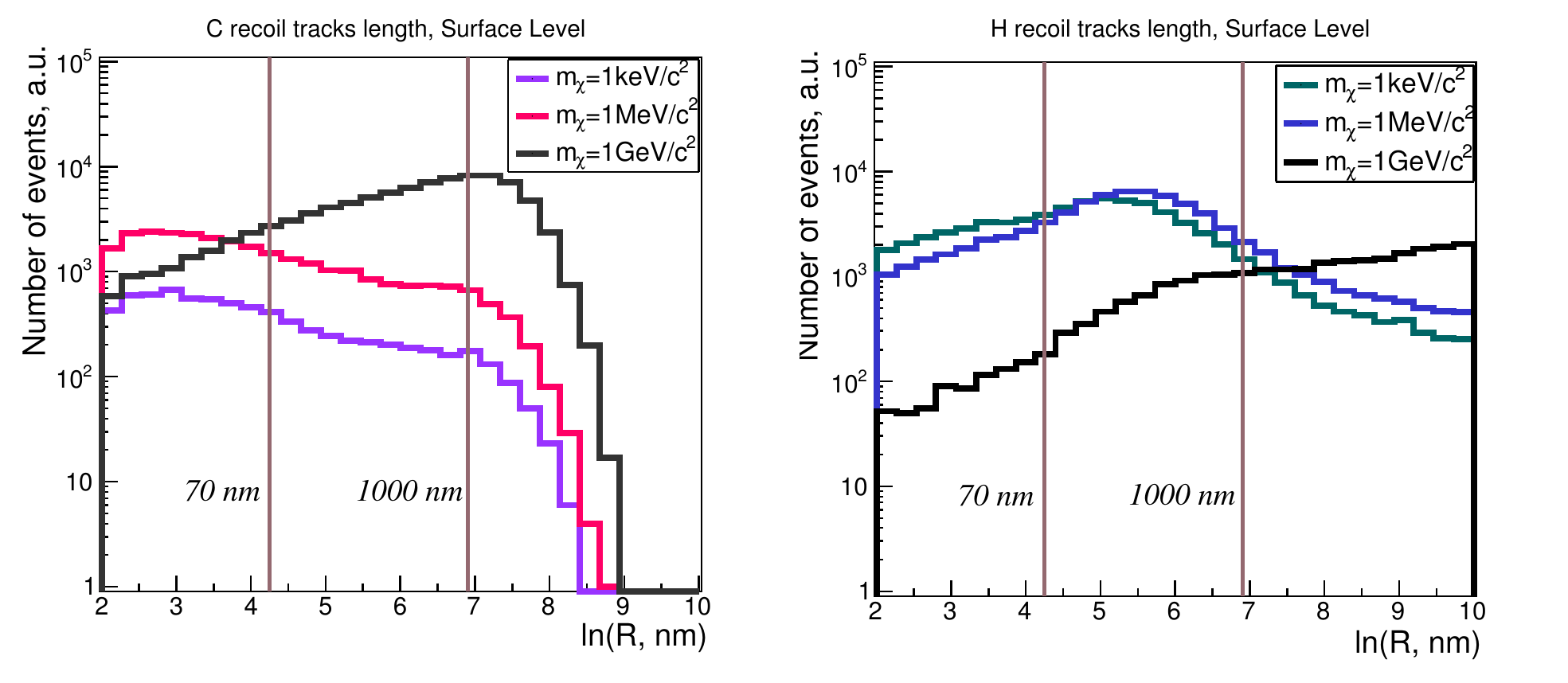} 
\caption{C (left) and H(right) recoil track length R distributions at the surface level. For $\sigma_{\chi p}=10^{-28} $~cm$^2$, $m_{\chi}$ = 1 keV/c$^2$, 1 MeV/c$^2$, 1 GeV/c$^2$ and for $10^5$ DM at the edge of the atmosphere. The vertical lines indicate the limits on the track lengths of 70 and 1000 nm.} 
\label{pic6}
\end{figure}

In figure~\ref{pic6}, the track length (ln R) distributions of carbon and hydrogen recoils at the surface laboratory are reported for three mass values and a cross-section $\sigma_{\chi p} = 10^{-28}$~cm$^2$. The GEANT4 tracking parameter $setCut$ for carbon nuclei was 0.1 mm, for hydrogen nuclei --- 10 nm. Hydrogen recoils for DM particles of all the considered masses have energies up to tens of MeVs. Consequently, the corresponding path lengths extend up to several centimeters. However, this has to be convoluted with the H mass fraction in the emulsion which is 1.63\%.  The distributions in Figure~\ref{pic6} extend to much longer lengths than in the non-relativistic dark matter case~\cite{A-wimp}. It is worth noting that most of the recoils exceed the track length of 70~nm. 

The fraction of events with track lengths above the 70~nm threshold is defined as the $k_{70}$ coefficient. Sensitivity curves are reported with track length thresholds of 70~nm and 1~$\mu$m, corresponding to the coefficients $k_{70}$ and $k_{1000}$.

\section{Nuclear recoil distributions and directional search}
\label{sec:nuclear_angle}

In order to evaluate the capability of a directional observation with NIT emulsion films, one has to first evaluate whether nuclear recoil tracks preserve the direction to the source of the scattering DM particles and how the directionality of the signal depends on DM particle mass, recoiling nucleus type, and observation level.

Secondly, one needs to estimate the difference between the primary DM particle fluxes coming from different directions, for instance, between the Galactic Center and that Galactic Anticenter or between the Galactic Center and the direction perpendicular to it.

Let us consider the former part first.

\begin{figure}[h]
\centering
\includegraphics[width=0.9\textwidth]{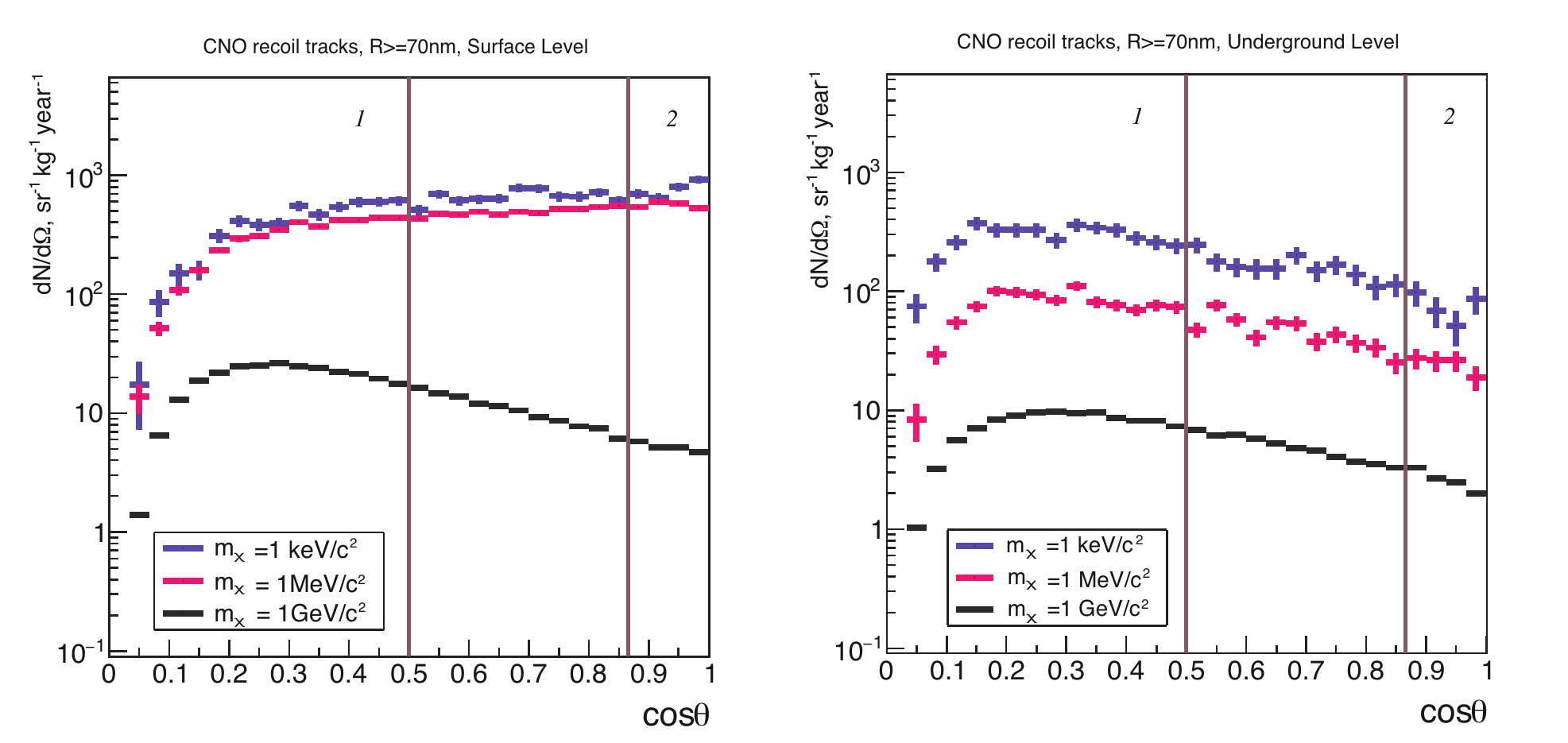} 
\caption{Angular distributions of C recoil events in sr$^{-1}$kg$^{-1}$year$^{-1}$ with track lengths $> 70$ nm at the surface and underground levels for $m_{\chi}$ = 1 keV/c$^2$, 1 MeV/c$^2$ and 1 GeV/c$^2$ as examples. Vertical lines denote the angular ranges [0, 30] --- 2 and [60, 90] --- 1.} 
\label{pic22}
\end{figure}

To determine how well the nuclear recoil track direction preserves the DM flux direction we 
compare the number of registered events in sr$^{-1}$ with angles between their directions and the direction to the source being in the ranges [0, 30] and [60, 90] degrees for various input parameters. Let their ratio be the \textit{directionality index} $k_{\frac{0-30}{60-90}}$.

\begin{figure}[h]
 \centering
\includegraphics[width=0.9\textwidth]{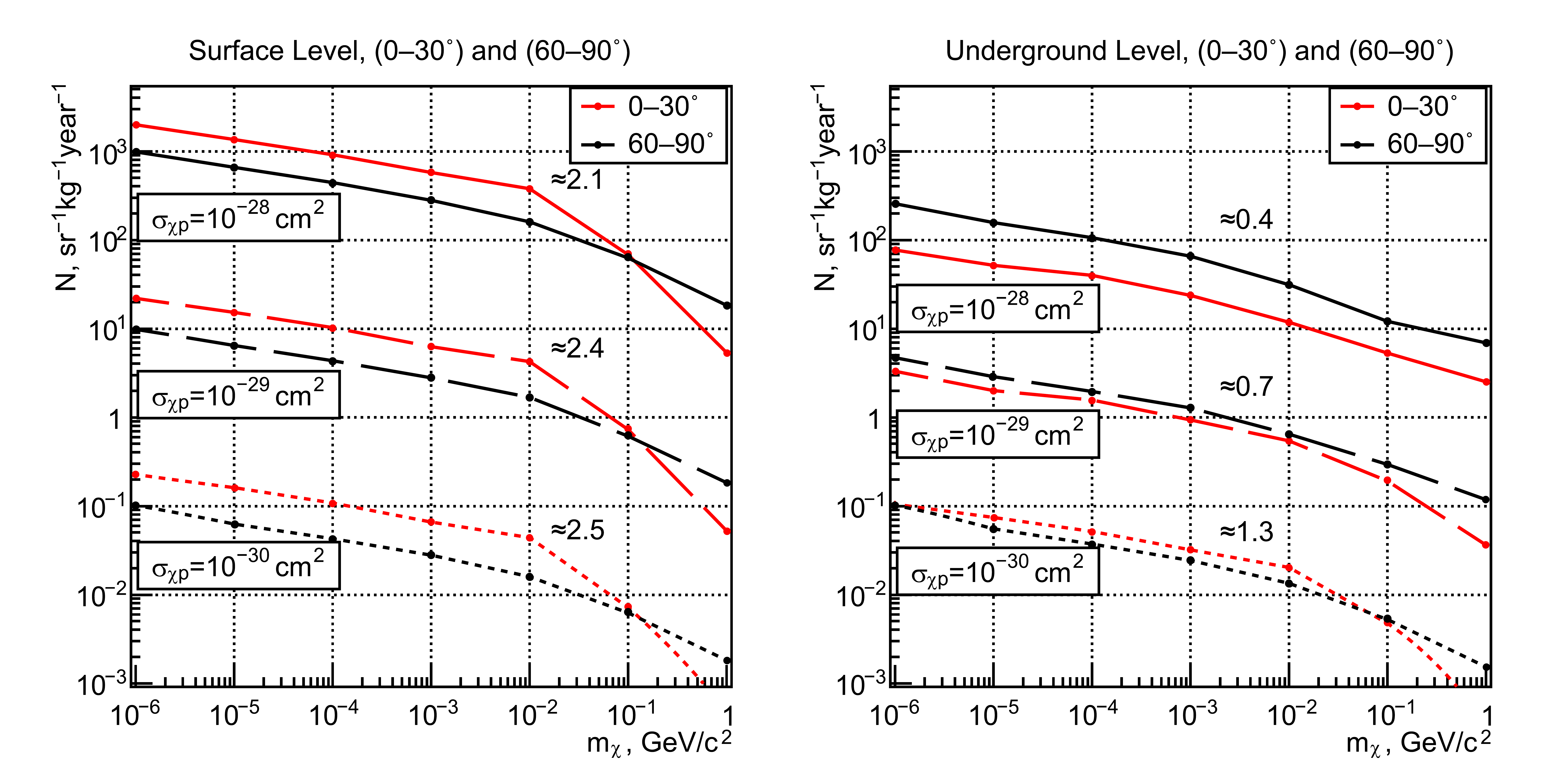} 
 \caption{Directionality effect for H and CNO recoil track events at the surface level (left) and underground level (right). Coefficients $k_{\frac{0-30}{60-90}}$ are given for DM mass ranges 1 keV/c$^2$ --- 10 MeV/c$^2$.} 
 \label{pic8dir}
\end{figure}

 The direction of a nuclear recoil produced by the DM is estimated by a fit to the track grains.
 To estimate the effect of directionality, one needs to compare the number of H and CNO recoil tracks for 10-kg NIT films, in the above-mentioned angular ranges, i.e. [0, 30] and [60, 90] degrees. To this end, we account for the material in 10 kg of emulsion, normalize the DM flux at the edge of the atmosphere and multiply it by the coefficient $k_{70}$ 
 denoting the ratio between the number of nuclear recoil tracks above the 70~nm threshold and the number of primary DM particles. The angular distributions of carbon nuclei tracks over a one-year exposure of 10 kg of emulsion are reported in figure~\ref{pic22}. Vertical lines denote the angular ranges for which the fluxes are compared.

Figure~\ref{pic8dir} shows the correlation between the direction of nuclear recoil tracks with primary DM particles at the surface and underground levels for three cross-section options $\sigma_{\chi p} = 10^{-30}, 10^{-29}$ and $10^{-28}$~cm$^2$. These figures also show the coefficients evaluating the directionality effect
$k_{\frac{0-30}{60-90}}$.

Figure~\ref{pic8dir} shows that at the surface level the signal in the [0, 30] degree range is more than twice as high as the one in the [60, 90] degree range. Moreover, the directionality analysis should account for different DM particle mass ranges, (1 keV/c$^2$ --- 10 MeV/c$^2$) and (100 MeV/c$^2$ --- 1 GeV/c$^2$). Indeed, figure~\ref{pic8dir} shows that the directionality effect is inverted at the surface for DM masses above 100 MeV/c$^2$: the number of recoil tracks in the [60, 90] degree range exceeds the one in the [0, 30] degree range. The panel on the right shows that at the underground level the directionality effect decreases substantially: the $k_{\frac{0-30}{60-90}}$ coefficient is of the order of 1 and even lower. This fact additionally supports the necessity to deploy the experimental apparatus at the surface.

\section{Results of a directional search}
\label{sec:directional}

\begin{figure}[h]
\centering
\includegraphics[width=0.55\textwidth]{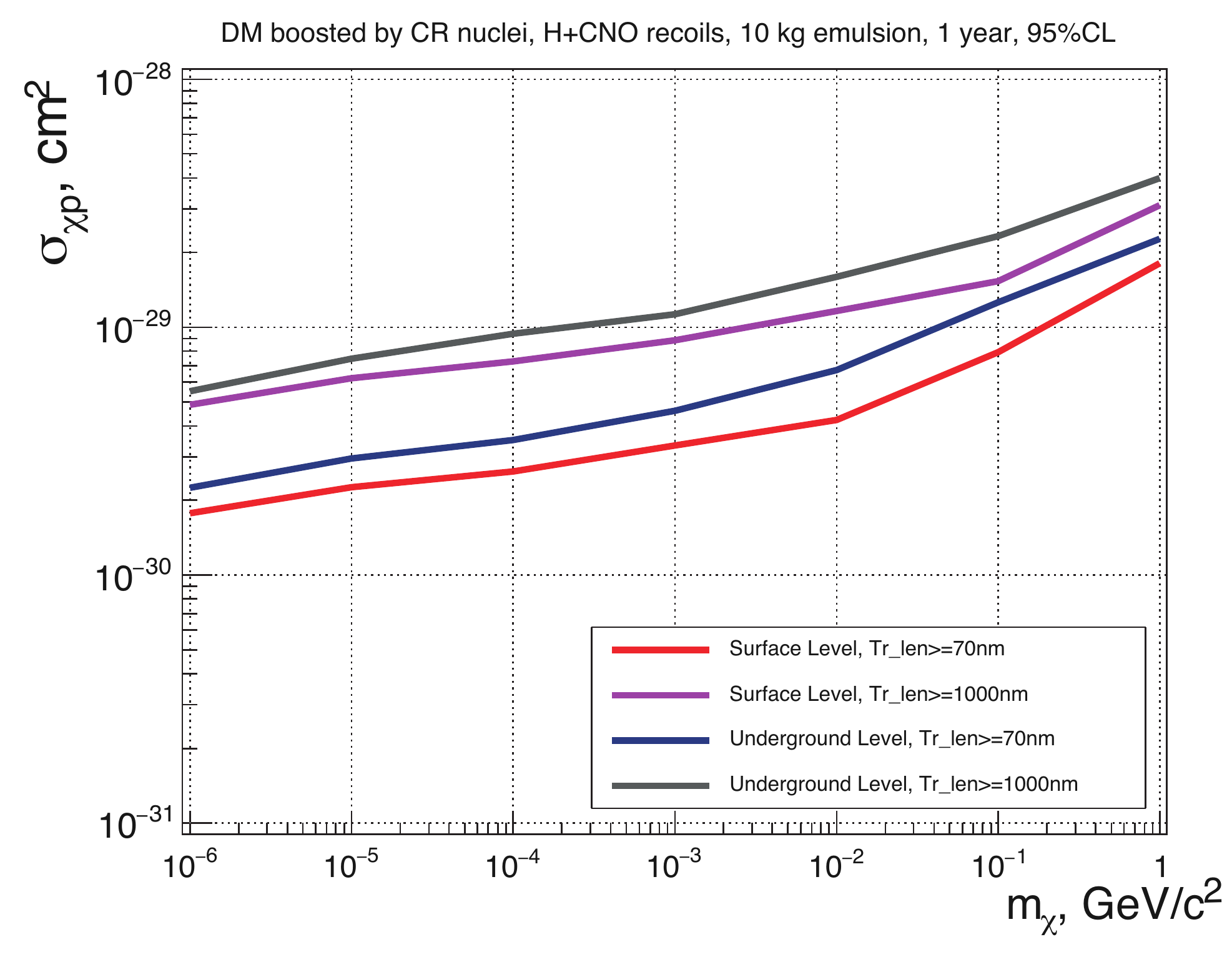}
\caption{Sensitivity curves of the NEWSdm detector in the form of 10 kg of nuclear emulsion with $\rho=$3.44 \; g/cm$^{3}$ for 1 year of exposure. The lines correspond to three CNO nucleus and hydrogen H recoil events with track length thresholds ($Tr\_len$) 70 and 1000 nm.} 
\label{pic15-1}
\end{figure}

To estimate the sensitivity of the emulsion detector to CRDM, we have modeled the fluxes of DM for all considered masses, cross sections $\sigma_{\chi p}$, at the surface and underground levels of observation. As a next step, we have calculated the number of hydrogen and carbon target nuclei in 10 kg of emulsion. The contribution of oxygen and nitrogen nuclei was accounted for by multiplying the carbon nuclei contribution by a factor of two.
As a final step, we have taken the cross section $\sigma_{\chi p}$ of DM interaction with target emulsion nuclei, and the coefficients $k_{70}$ and $k_{1000}$ into account for the thresholds. The sensitivity curves have been obtained via simple interpolation methods for intermediate cross-section values $\sigma_{\chi p}$. 
The sensitivity curves shown in figure~\ref{pic15-1} account only for the contribution from the vicinity of the center of the Galaxy within $\pm5$ degrees. As shown by subsequent calculations, integrating over the central hemisphere inside the Galactic Disk improves the sensitivity of NEWSdm detector by a factor of 2 to 3.

\begin{figure}[h]
    \centering
  \includegraphics[width=0.6\textwidth] {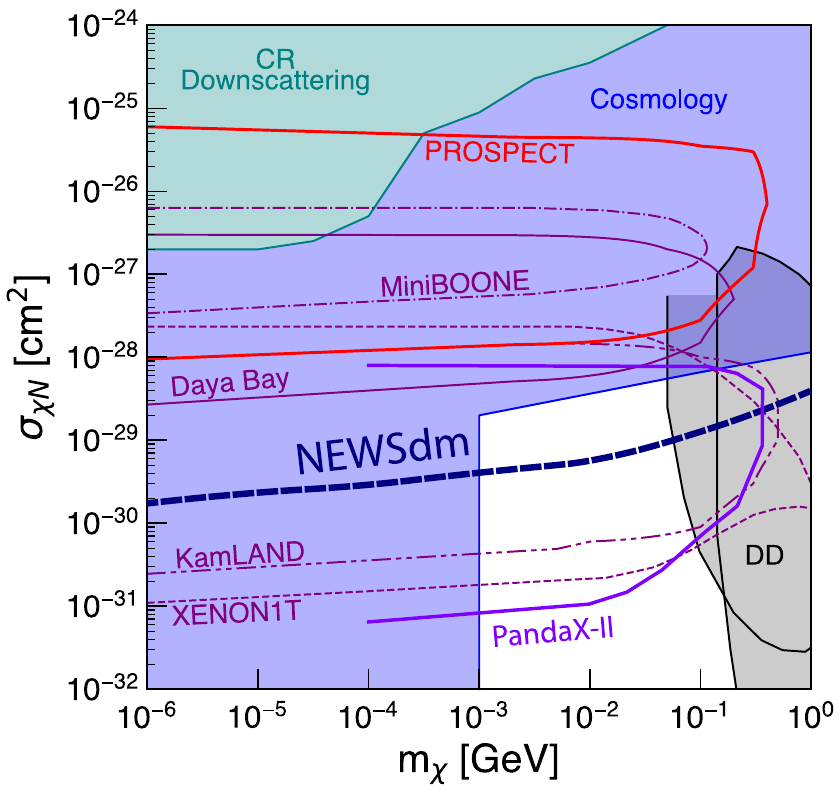} 
      \caption{Sensitivity curves of the NEWSdm detector with 10 kg of nuclear emulsion for 1 year of exposure at the surface level superimposed to limits derived from Refs.~\cite{Pospelov2019},~\cite{Cappiello2},~\cite{Cappiello2-Erratum} and ~\cite{PROSPECT}. The boundaries go through the dots corresponding to three H and CNO recoil events with track lengths longer than 70 nm for zero background.} 
    \label{pic15-3}
\end{figure}

The sensitivity curve for a 10 kg year exposure of the NEWSdm experiment at the surface level with the 70~nm threshold is drawn on Figure~\ref{pic15-3} together with the exclusion contours from cosmology, underground direct detection experiments and neutrino experiments derived from Refs.~\cite{Pospelov2019},~\cite{Cappiello2},~\cite{Cappiello2-Erratum}.
We discuss here the experimental constraints on Figure~\ref{pic15-3}. 

At first, it should be pointed out that most of the experimental limits drawn in this Figure are obtained by groups~\cite{Pospelov2019},~\cite{Cappiello2},~\cite{Cappiello2-Erratum} who have used data published by the different Collaborations and recomputed the sensitivity. This is the case for the XENON1T, KamLAND, Daya Bay and MiniBOONE experiments. Only PROSPECT~\cite{PROSPECT} and PandaX-II~\cite{atten-PANDA} Collaborations have done their own analysis. 

XENON1T, PandaX-II, Kamland and Daya Bay are experiments located deep underground. Indeed, XENON1T is a Xe TPC detector located at the INFN Gran Sasso laboratory, under 1400 m of rock. PandaX-II~\cite{atten-PANDA} is a Xe TPC detector located under 2400 m of rock. Kamland is an isoparaffine based  liquid scintillator detector, more than 1000 m deep underground. Daya Bay is a liquid scintillator, at a depth ranging from 250 to 860 water equivalent meters. Being underground introduces significant distortions in the energy spectrum and makes the result less model independent. 

The limits on $\sigma_{\chi p}$ from XENON1T were obtained in Ref.~\cite{Pospelov2019} by recalculating the sensitivity obtained for DM masses above 100 GeV/$c^2$ ~\cite{Pospelov2019-Xenon} and porting it to the sub-GeV DM  masses. It is worth noting that the XENON1T collaboration has published a sub-GeV Dark matter search based on the analysis of recoil electrons signal~\cite{Xenon-electron}.

Figure~\ref{pic15-3} shows the  95\% CL exclusion limit from  the PROSPECT~\cite{PROSPECT} experiment. The apparatus consists of a $^6$Li-doped liquid scintillator detector operated  on the surface. MiniBOONE is a Cherenkov-based neutrino detector located at the Fermilab US Laboratory and operated near the surface. The configuration of both PROSPECT and MiniBOONE are similar to the one of the NEWSdm experiment from the point of view of operation conditions: they can be directly compared. 

Almost the entire range of sub-GeV DM masses  for  $\sigma_{\chi p}> 10^{-33}$ $cm^2$  is excluded in ~\cite{atten-S-K}  based  on the SuperKamiokande(SK) Cherenkov underground detector data for two different cross section models. However, the threshold of proton Cherenkov radiation for SK  is reported to be above 1 GeV~\cite{atten-S-K}. Therefore, as we have shown in this work, such energetic recoil protons will have a recoil direction  perpendicular to the DM  momentum direction, and only the tail of the distribution of boosted DM scattering proton events can be detected with the cut they used on the proton momentum.

Despite the fact that the range of cross sections $\sigma_{\chi p} = 10^{-30} - 10^{-28}$ $cm^2$ for DM masses from 1 keV/c$^2$ to 1 GeV/c$^2$ have been already investigated via different data samples, the location of the emulsion detector on the Earth’s surface with its capability to detect the direction will bring additional insight in the overall picture. This is particularly true since the most strongest experimental limits come from underground experiments. Indeed,  the propagation of dark matter particles through a large rock thickness can distort the spectrum and introduce additional uncertainties, thus biasing the results. This strengthens the interest in an emulsion-based  detector installed on the surface, with high-resolution tracking capabilities.



In section~\ref{sec:nuclear_angle}, we have shown that the direction to the GC can be identified when analyzing the directions of nuclear recoil tracks produced by DM particles accelerated in elastic interactions with cosmic rays. It is assumed that the concentration of both DM and CR is highest near the Galactic Center. Similar estimates have been presented in~\cite{Sup-Hip} for Super-Kamiokande and Hyper-Kamiokande underground experiments.

\begin{figure}[h]
\centering
\includegraphics[width=0.45\textwidth]{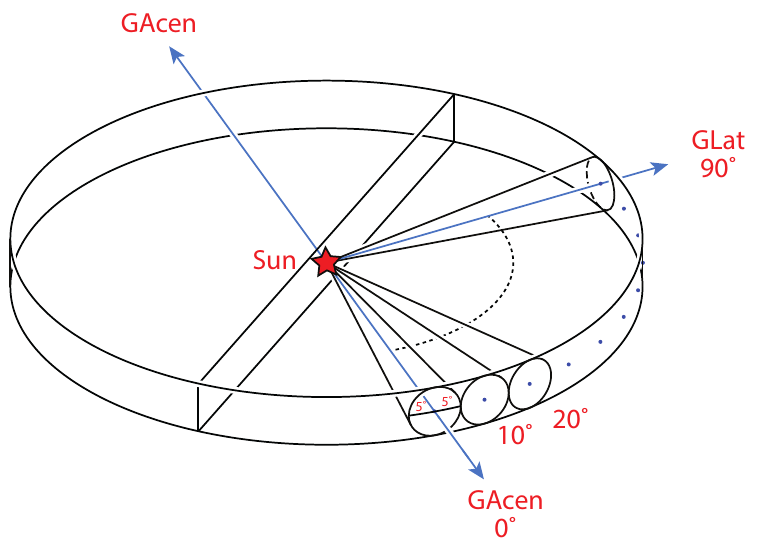}
\caption{CRDM flux calculation scheme from the GC to GLat directions.} 
\label{schemeVasilisa}
\end{figure}

We now estimate the ratio of the number of nuclear recoils from DM particles (for the adopted CRDM model with $\sigma_{\chi p}=10^{-28}$~cm$^2$ as an example) coming from the GC direction to the one from DM particles arriving from the perpendicular (GLat) direction. Based on the CR and DM distribution maps shown in figure~\ref{pic2}, one can assume that the signal of GLat direction will be significantly different from the GC signal.

To be able to compare the number of recoil tracks from GC and GLat, we evaluated the CRDM fluxes for the full set of $\pm 5$ degree cones within 0--90 degrees with respect to the GC direction (i.e. from GC to GLat within the plane of the Galaxy), see figure~\ref{schemeVasilisa}. 

\begin{figure}[h]
\centering
\includegraphics[width=0.45\textwidth]{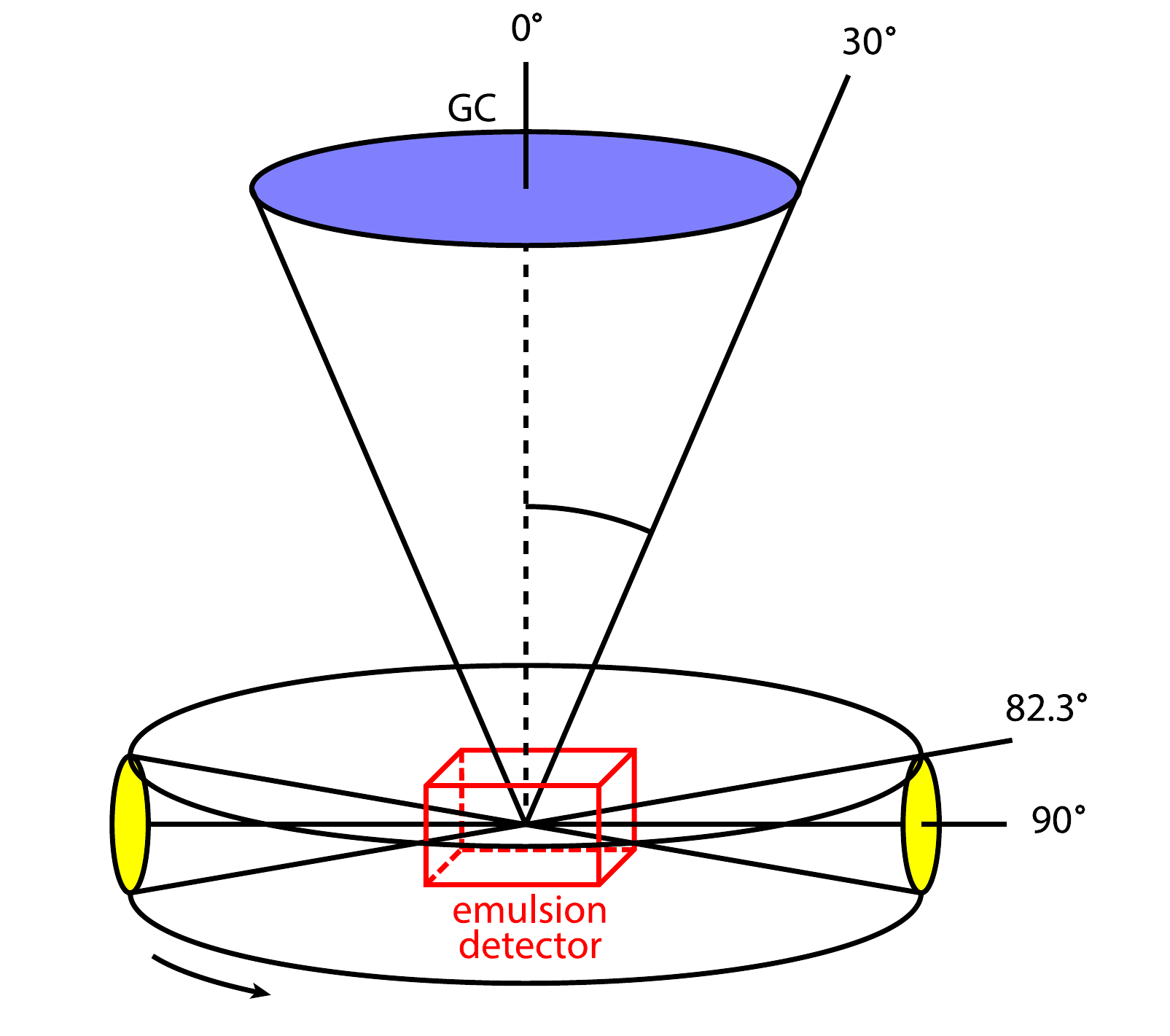}
\caption{The directional signal evaluation scheme. The solid angle indicated by the blue marker corresponds to the Galactic Center direction, the yellow marker denotes an equal solid angle in the perpendicular direction.} 
\label{pic11-sc}
\end{figure}

We have obtained a continuous set of fluxes of CRDM in cones centered at 0, 10, 20, 30, 40, 50, 60, 70, 80, 90 degrees. For 0, 10, 45, 90 degrees centered cones we used the full calculation scheme, the fluxes from other cones were evaluated by approximation. From this, we now have continuous CRDM flux from GC to GLat. Now we can evaluate the angular distribution of the nuclear recoil tracks for every CRDM source, i.e. the cones contributing to the signal. At the last stage, we have to sum up all the recoils' angular distributions coming from all the cones and define if the signal excess is observable, according to the considered model. 

Figure~\ref{pic11-sc} shows the emulsion detector and equal solid angles in the direction of the supposed signal source (Galactic Center) and in the perpendicular direction within which the number of recoil tracks was compared. This proves that the emulsion detector should be located on the equatorial telescope, which will make it possible to keep a constant orientation of the emulsion plates relative to the signal source.

\begin{figure}[h]
\centering
\includegraphics[width=1.0\textwidth]{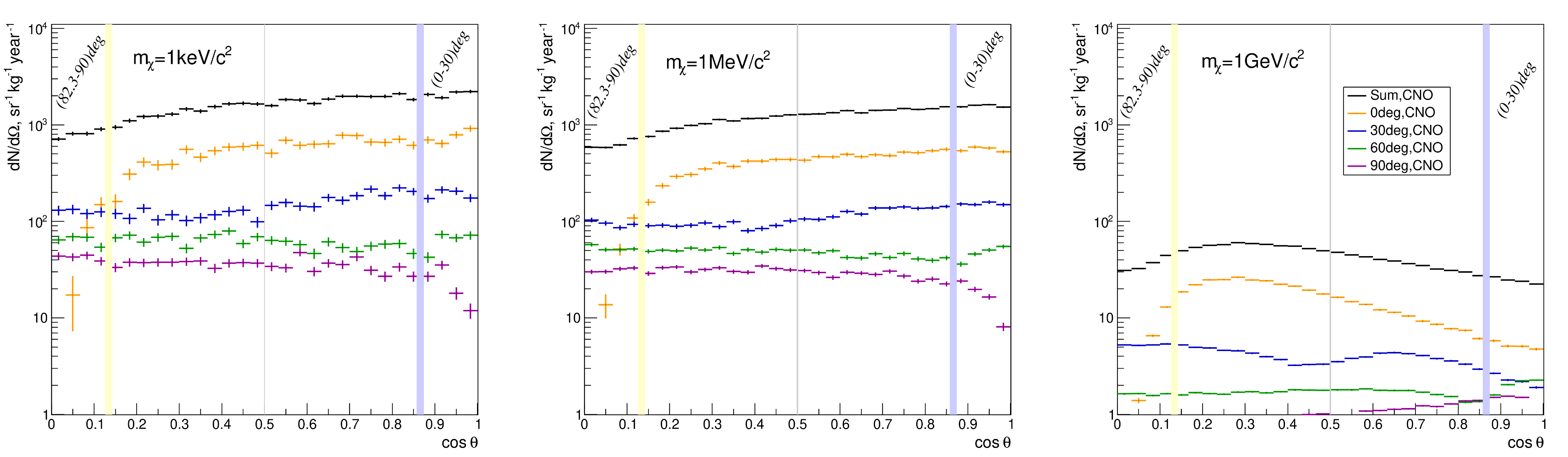}
\caption{CNO nuclear-recoil angular distributions from boosted DM particles with $m_{\chi}$ = 1 keV/c$^2$ (left), $m_{\chi}$ = 1 MeV/c$^2$ (center) and $m_{\chi}$ = 1 GeV/c$^2$ (right) coming from GC to the perpendicular direction. Colored lines stand for selected angular distributions of the CNO recoils from DM with $0,\; 30,\; 60,\; 90$ degree cone directions. Black histograms represent the total CNO angular distributions. $\sigma_{\chi p} = 10^{-28} $~cm$^2$.} 
\label{pic12}
\end{figure}

Histogram samples shown in figure~\ref{pic12} make it possible to compare the angular distributions of nuclear recoils from DM particles of different masses.
Value $\cos\theta = 1$ corresponds to the GC (expected DM signal) direction, $\cos\theta = 0$ --- to the perpendicular one. The colored lines stand for the angular distributions of the CNO recoils caused by the CRDM coming from some selected directions with respect to the GC. The black histograms show the total CNO angular distribution. Naturally, the angular distributions of hydrogen recoils were also taken into account, but because of the small contribution, they are not shown in the figures.

The vertical yellow and blue lines in figure~\ref{pic12} highlight half of the opening angle of the cone, corresponding to equal solid angles near the GC and perpendicular directions, see figure~\ref{pic11-sc}. The directionality effect can be estimated by comparing the number of nuclear recoil tracks in these ranges. The results are shown in Table 1. The numbers of nuclear recoils tracks in sr$^{-1}$kg$^{-1}$year$^{-1}$ in the first four (0--3) $N_\text{0--3}$ and the last four (26--29) $N_\text{26--29}$ histogram bins are compared.

\begin{table}[h]
\begin{center}
\caption{Directivity parameters}
\begin{tabular}{|l|c|c|c|}
\hline
DM mass & \multicolumn{1}{l|}{$N_{0-3}$} & \multicolumn{1}{l|}{$N_{26-29}$} & \multicolumn{1}{l|}{$\frac{N_{26-29}}{N_{0-3}}$} \\ \hline
1 keV/c$^2$ & 2319.8 & 8158.5 & 3.5 \\
1 MeV/c$^2$ & 1820.2 & 6531.2 & 3.6 \\
1 GeV/c$^2$ & 100.9 & 102.4 & 1.0 \\
 \hline
\end{tabular}
\end{center}
\end{table}

Based on figure~\ref{pic12} and Table 1, it can be concluded
that 3.5 times more tracks of CNO and H recoil nuclei in the emulsion can be expected from DM particles with masses from 1 keV/c$^2$ to 1 MeV/c$^2$ coming from the Center of the Galaxy than from those coming from the perpendicular direction. Taking the directivity effect illustrated in Fig. 8 into account, it can be assumed that the search for a directed signal is possible only on the surface and for DM particles with masses up to several tens of MeV.


\section{Conclusions}
\label{one1}
We have reported a study on the nuclear recoils induced by CRDM interactions with the nuclei of the
NIT emulsion films for a module of the NEWSdm detector. From the background point of view, the search for CRDM can be carried out both underground and on the surface, given the relatively long track lengths expected for nuclear recoils induced by a CRDM particle. Therefore, we have studied two scenarios for the INFN Gran Sasso Laboratory: installation at the surface level or in the underground laboratory. 

At the surface laboratory, one can expect a factor of 3.5 in the ratio between the number of recoil track events detected in the direction of the Galactic Center when compared to the orthogonal direction. On the contrary, given the attenuation in the rock, the correlation of the measured nuclear recoil directions with the incoming flux gets lost underground to a large extent. Indeed, the relatively large cross-section induces a significant attenuation and scattering of CRDM particles in the thick rock layer. This effect depends on the DM mass and on the interaction process, thus introducing a sizeable bias and deterioration for directional observation. 

It follows that the module of the NEWSdm detector is to be preferably placed at the surface laboratory for a high-sensitivity directional CRDM search, due to smaller systematic uncertainties and for a better directionality preservation.

A module of the NEWSdm apparatus consisting of 10 kg emulsion detector exposed for one year at the surface laboratory on an equatorial telescope can independently explore the existence of cosmic ray boosted DM particles in the mass range from 1 keV/c$^2$ to 1 GeV/c$^2$ for cross-section values down to 10$^{-30}$~cm$^2$. 


\section*{Acknowledgements}
This work is partially supported by a RSF 23-12-00054 grant and Interdisciplinary
Scientific and Educational MSU School ``Fundamental and Applied Space Research''.





\end{document}